\documentclass[twocolumn,showpacs,preprintnumbers,amsmath,amssymb,prl,superscriptaddress]{revtex4}

\usepackage{graphicx}
\usepackage{color}
\usepackage{epsfig}
\usepackage{overpic}

\newcommand{\vb}[1]{{\boldsymbol{#1}}}

\begin{document}
 
\title{Semiclassical polaron dynamics of impurities in ultracold gases}
\author{David Dasenbrook}
\author{Andreas Komnik}
\affiliation{Institut f\"ur Theoretische Physik, Universit\"at Heidelberg, Philosophenweg 19, D-69120 Heidelberg, Germany}

\date{\today}

\begin{abstract}

  We present a semiclassical treatment of a fermionic impurity coupled to Bogolyubov modes of a
  BEC. In the lowest order approximation we find a full solution of an initial value problem, which
  turns out to behave differently in the sub- and supersonic regimes. While in the former case no
  impurity deceleration is observed, in the latter case non-Markovian dissipation effects kick in
  resulting in slowing down of the fermion. Although this scenario is compatible with the one
  offered by an elementary field theoretical picture at weak coupling, the details of the dynamics
  turn out to be completely different. Fluctuation effects can be taken into account by expanding
  around the classical path, which leads to a natural cutoff scale for the momentum integrals of the
  order of the inverse polaron radius. As an application we calculate the drag force which is
  exerted by the BEC on the impurity moving with constant velocity $v$. Contrary to the perturbative
  result, according to which the drag force is $\sim v^4$, it turns out to be proportional to
  $1/v^2$ in the semiclassical regime.

\end{abstract}
\pacs{03.75.Mn, 71.38.Fp, 67.85.Pq, 78.20.Bh}

\maketitle

%\section{Introduction}
\label{sec:intro}

Interacting ultracold gases offer unique opportunities for the modeling of realistic quantum systems
\cite{RevModPhys.80.885}. During the recent years it not only became possible to study such exotic
phenomena as BEC-BCS crossover, but also the modeling of systems inspired by the fractional quantum
Hall effect might soon be within reach \cite{Chen20051,Bloch:2012fk,Gurarie20072}.  One of very
interesting and old condensed matter physics phenomena, which shows exciting physics on the one hand
and on the other hand might immensely profit from parameter adjustability offered by ultracold gas
systems is the polaron problem \cite{feynman1998statistical}. In its simplest realization it
describes electrons in a semiconductor which interact with phonons of the background lattice. The
most important feature of the model is a fundamental difference in the properties of the system for
weak and strong coupling between the fermionic and bosonic subsystem.

It turns out that polaron-like models can very effectively be simulated by fermionic impurity atoms
immersed into a BEC \cite{PhysRevA.76.011605,PhysRevB.80.184504}. Alternatively one can couple them
to a continuum of bosonic excitations of some other fermionic species
\cite{PhysRevLett.102.230402}, or even use bosonic impurities of the species different from the one constituting the BEC bath. A direct mapping of most of the results from the vintage literature
is, however, not possible for a number of reasons. First of all, the spectrum of bosonic excitations
is usually profoundly different from those in a semiconductor. The second distinction is the
coupling mechanism between the subsystems. Some less important details such as the presence and
nature of cut-off parameters might drastically alter the properties of a given system as well. But
the most important difference is at the same time an enormous advantage -- it is the adjustability
of parameters, which is far superior to that in solid state realizations. This feature makes it
possible to advance into new and as yet unexplored parameter regions. One such situation is the
semiclassical limit, in which the impurity can be considered to be an almost classical particle. Not
only is this particular parameter constellation so far poorly understood, it also allows for a
physically intuitive characterization of dynamics in super- and subsonic velocity regimes that have
mostly been investigated using involved numerical methods in similar systems before
\cite{Mathy:2012fk,PhysRevLett.97.260403}.

We consider a single fermionic impurity interacting with a gas of bosons with masses $m$ (we use
polaronic units, where the mass of the impurity $m_I$ as well as the healing length of the BEC $\xi$
and the reduced Planck constant $\hbar$ are set to unity, throughout the Letter). The Hamiltonian
reads
\begin{align}
  \label{eq:hamiltonian1}
  H = \frac{\vb{p}^2}{2} + \sum_\vb{k} \epsilon_\vb{k} a_\vb{k}^\dagger a_\vb{k} &+ \frac{1}{2}
  \sum_{\vb{k},\vb{k}',\vb{q}} V_{BB}(\vb{q}) a_{\vb{k}'-\vb{q}}^\dagger a_{\vb{k}+\vb{q}}^\dagger
  a_\vb{k} a_{\vb{k}'} \nonumber\\
  &+ \sum_{\vb{k}',\vb{q}} V_{IB}(\vb{q}) \rho(\vb{q}) a_{\vb{k}'-\vb{q}}^\dagger a_{\vb{k}'},
\end{align}
where $a_\vb{k}^\dagger$ and $a_\vb{k}$ are the creation and annihilation operators for a boson with
momentum $\vb{k}$, $\vb{p}$ is the momentum operator of the impurity, $V_{BB}$ and $V_{IB}$ are
general boson-boson and impurity-boson interaction potentials, and the dispersion of the bosons is
given by
\begin{equation}
  \label{eq:dispersionbosons}
  \epsilon_\vb{k} = \frac{\vb{k}^2}{2m} - \mu \, .
\end{equation}
$\mu$ is the chemical potential. The momentum space density of a point particle is related to
its position operator $\vb{x}$ via
\begin{equation}
  \label{eq:momentumspacedensitypointparticle}
  \rho(\vb{q}) = \int \mathrm{d}^3 y \delta^3(\vb{y}-\vb{x}) e^{i \vb{q} \vb{y}}.
\end{equation}
Since we assume a dilute homogeneous gas, we perform the Bogoliubov approximation by
replacing the zero mode operators $a_0$, $a_0^\dagger$ by $c$-numbers \cite{pitaevskij2000bose}. 
We furthermore use contact pseudo-potentials for the
interactions $V_{BB}(\vb{q}) = g_{BB}$, $V_{IB}(\vb{q}) = g_{IB}$, where the constants $g_{BB}$ and
$g_{IB}$ are related to the $s$-wave scattering lengths of the true potentials via the
Lippman-Schwinger equation. We then obtain a variant of the Fr\"ohlich Hamiltonian
\cite{AdvPhys.3.325,PhysRevB.80.184504}
\begin{equation} 
  \label{eq:hamiltonian2}
  H = \frac{\vb{p}^2}{2}  + \sum_\vb{k} \epsilon_\vb{k} b_\vb{k}^\dagger
  b_\vb{k} + \sum_\vb{k} V_\vb{k} e^{i \vb{k} \vb{x}} \left( b_\vb{k} + b_{-\vb{k}}^\dagger \right).
\end{equation}
For reasons that will become clear below, we restrict the dispersion relation of the Bogoliubov
quasiparticles $\epsilon_\vb{k}$ and the interaction structure $V_\vb{k}$ to the expressions valid
for momenta that are smaller than the healing length in the BEC. 
Therefore, we have
\begin{equation}
  \label{eq:epsilonk}
  \epsilon_\vb{k} = c k, \qquad V_\vb{k} = \sqrt{\alpha k},
\end{equation}
where $c = 1/(\sqrt{2} m)$ is the sound velocity in the BEC and the dimensionless coupling
constant $\alpha$ can be related to the density $n$ of the bosonic gas and the impurity-boson
interaction constant via $\alpha = n g_{IB}^2 / \sqrt{2}$.

Using the standard procedures, we can now integrate out the environmental (BEC) degrees of freedom,
see for example~\cite{weiss1999quantum}. A requirement for a semiclassical treatment is that the average
wavelength of the impurity be small compared to a typical length scale of the system
\cite{messiah2000quantum}. The latter is given by the healing length $\xi$, since it is the typical
length on which the effective potential seen by the particle varies. The wave length of the impurity
particle is in a first approximation given by $\lambda = \hbar/(v m_I)$, where $v$ is its
velocity. Since the healing length is proportional to $\hbar/(c m)$, we see that a large
impurity mass compared to the masses of the bosonic particles justifies a semiclassical
treatment. Moreover, we will mostly be interested in impurity velocities $v \geq c$, so that, for instance,
a ${}^{85}\text{Rb}$ impurity in a gas of ${}^{23}\text{Na}$ should well be described by this
theory. For the case in which $m_I < m$ however, we still expect this approximation to be valid for very fast impurities with $v \ge m c/m_I$.  
Alternatively, in the strong coupling regime, where $\alpha \gtrsim 1$ the self-trapping
effects are expected to restrict the extent of the polaron wave function to a region smaller than
the healing length of the BEC \cite{PhysRevB.80.184504}, or equivalently, the strong coupling leads to
the generation of a large effective impurity mass. Apart from these effects, the dynamical aspects
of the physics can then still be captured by the classical equations of motion, provided one works
with the effective quantities.

\begin{figure}
  \centering
  \includegraphics[width=\columnwidth]{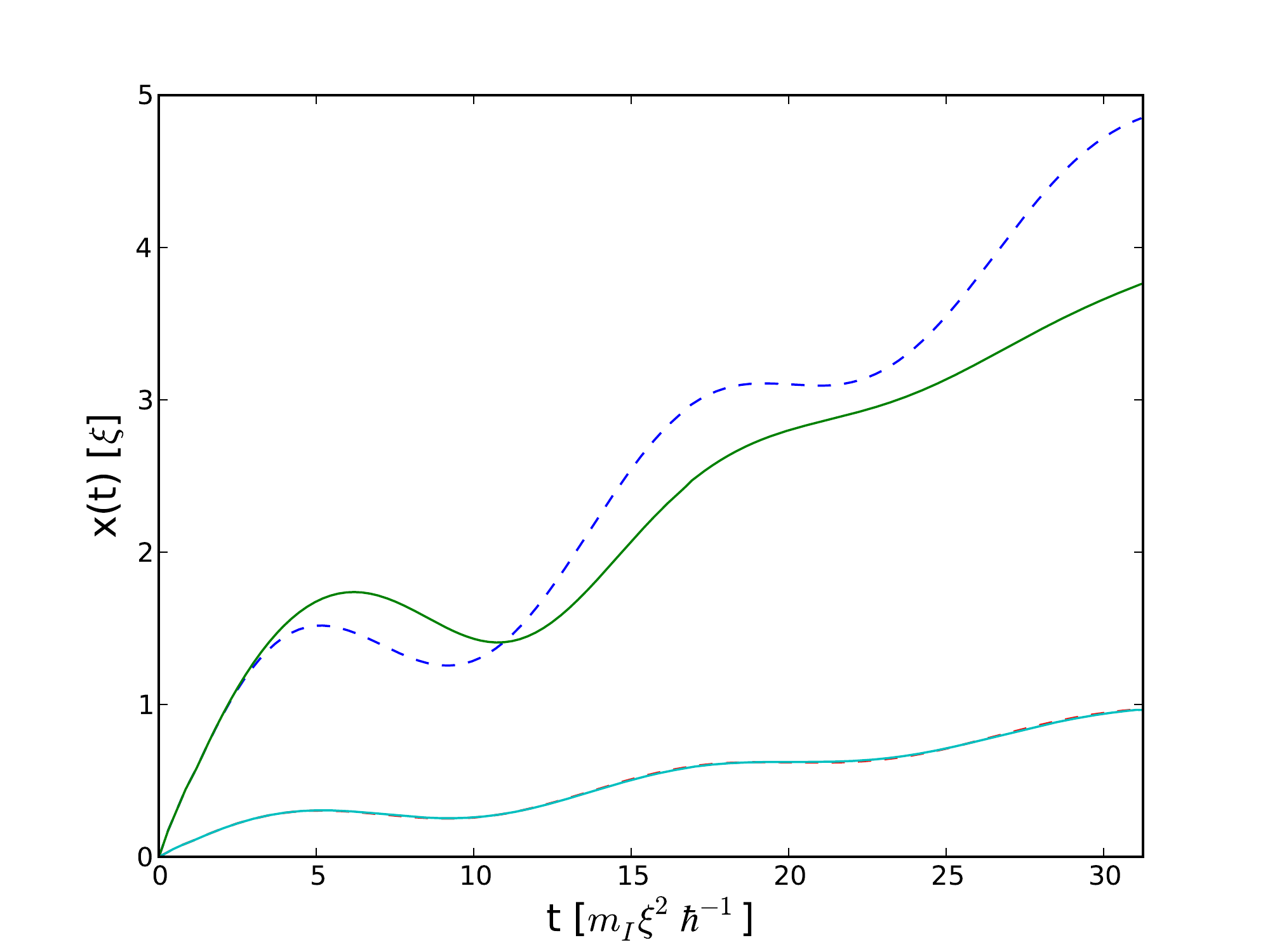}
  \caption{
 Results of the numerical solution of Eq.~\eqref{eq:eom} at short to intermediate timescales. The lower curves correspond to an initial condition $v_0= 0.5c$ whereas the upper curves represent the supersonic case $v_0= 3c$. Dashed lines are computed in the assumption ${\bf r}=0$, see text.
  }
  \label{fig:numericsolution}
\end{figure}

In order to preserve the convergence of the integrals within the employed approximation, it is
necessary to employ a cutoff scheme \cite{PhysRevB.32.3515}. For reasons which become clear below we
use a Gaussian $\sim e^{-k^2/k_c^2}$ where $k_c$ is proportional to the inverse of the size of the
polaron particle. The resulting EOM in three dimensions is then given by
\begin{equation}
  \label{eq:eom}
  \ddot{x}_i(t) = - \int_0^t \mathrm{d} u \gamma_{ij}(t-u,r) \dot{x}_j(u)
\end{equation}
with a generalized damping kernel
\begin{widetext}
\begin{align}
  \label{eq:generalizeddampingkernel}
  \gamma_{ij}(s,r) = &\delta_{ij} \frac{\alpha k_c^3}{16 \pi^{3/2} c r^3} e^{- \frac{1}{4} k_c^2 c^2
    s^2} \bigg\{ \left( e^{csk_c^2r} + 1 \right) \left(c^2 s^2 k_c^2 r + k_c^2 r^3 \right) -
  \left( e^{csk_c^2r} - 1 \right) cs \left( 1 + k_c^2 r^2 \right) \nonumber\\
  &\quad + \frac{r_j}{2r} \Big[ \left( e^{csk_c^2r} - 1 \right) \left( 3cs \left(4 + 2 k_c^2 r^2 +
      k_c^4 r^4 \right) - k_c^4 r^2 c^3 s^3 \right) - \left( e^{csk_c^2r} + 1 \right) \left( 3 k_c^2
    c^2 s^2 r \left( 2 + k_c^2 r^2 \right) + k_c^4 r^5 \right) \Big] \bigg\} \, , 
\end{align}
\end{widetext}
where ${\bf r = \vb{x}({\it t}) - \vb{x}({\it u})}$. The dependence of the rhs of Eq.~(\ref{eq:eom})
on $\bf r$ is a major structural difference to the case of linear coupling to the environment as in
the Caldeira-Leggett model \cite{PhysicaA.121.587}. This results in a dissipation dynamics which is
completely different. The influence of such non-Markovian dissipation scenario is weak for short
time scales and much of the dynamics can be described by an effective Caldeira-Leggett
model. However, the deviations are drastic for intermediate and long times. In this case the
solution of the initial value problem can only be found numerically \cite{masterthesis_text}. We
solve Eq.~\eqref{eq:eom} for the initial value $x(0)=0$ and $\dot{x}(0)=v_0$. As long as $v_0<c$
(subsonic initial velocity) the difference to a solution in which $r=0$ is set on the rhs of
Eq.~\eqref{eq:generalizeddampingkernel} throughout is minimal, see
Fig.~\ref{fig:numericsolution}. On the contrary, in the supersonic case $v_0 > c$ the non-Markovian
dissipation leads to drastic discrepancies.  In the transient regime we observe distinct damped
oscillations which can be connected with the appealing physical picture emerging in the elementary
self-consistent harmonic approximation \cite{feynman1998statistical}, according to which the polaron
cloud is modeled as an additional mass on a spring attached to the impurity. An important point is
the fact that both approximations cover the nonperturbative regime.
%%%%%%%%%%%%%%%%%%%%%%%%%%%%%%%%%
%%%%%%%%%%%%%%%%%%%%%%%%%%%%%%%%%

Despite its efficiency the numerical procedure is not able to produce reliable results for the
long-time asymptotics of the particle. In order to obtain it one has to rely on simplifications of
the basic equation \eqref{eq:eom}. As we have shown above, the dependence of the rhs on the details
of the particle trajectory at previous times via ${\bf r}$ is an obstacle. However, if we are
interested in the long-time limit, we can neglect fast fluctuations of the particle velocity, as
seen in the numerical simulations. Thus we can expand ${\bf r}(t,u) = {\bf v}(t)(t-u) + \dots$ and
require ${\bf v}(t)$ to be self-consistently calculated. For the long-time asymptotics one then
obtains the following EOM for the component of the displacement along the initial velocity
$\vb{v}_0$:
\begin{equation}
  \label{eq:longtimeeom}
  \lim_{t \to \infty} \ddot{x}(t) = \frac{\alpha}{8 \pi} \frac{c k_c^4 (c - \dot{x}(t) - |c -
    \dot{x}_i(t)|)}{(c - \dot{x}(t)) \dot{x}(t)^2}
\end{equation}
It is important to note that in this effective equation for the long time behaviour of the particle
coordinate, memory effects have not been neglected, but incorporated into the $t \to \infty$ limit.
The rhs is zero for $v(t)<c$ and $\propto v(t)^{-2}$ otherwise. This means that the solution to the
initial value problem $x(0)=0$, $\dot{x}(0) = v_0$ with Eq.~(\ref{eq:longtimeeom}) is
\begin{equation}
  \label{eq:longtimesolution}
  x(t) = 
  \begin{cases}
    v_0 t & \text{if } \dot{x}(t) < c \\
    \frac{2 \pi}{\alpha c k_c^4} \left( v_0^4 - \left|3 v_0^3 - \frac{3 \alpha c k_c^4}{8 \pi} t\right|^{4/3}
    \right) & \text{otherwise}
  \end{cases}
\end{equation}

\begin{figure}
  \centering
  \includegraphics[width=\columnwidth]{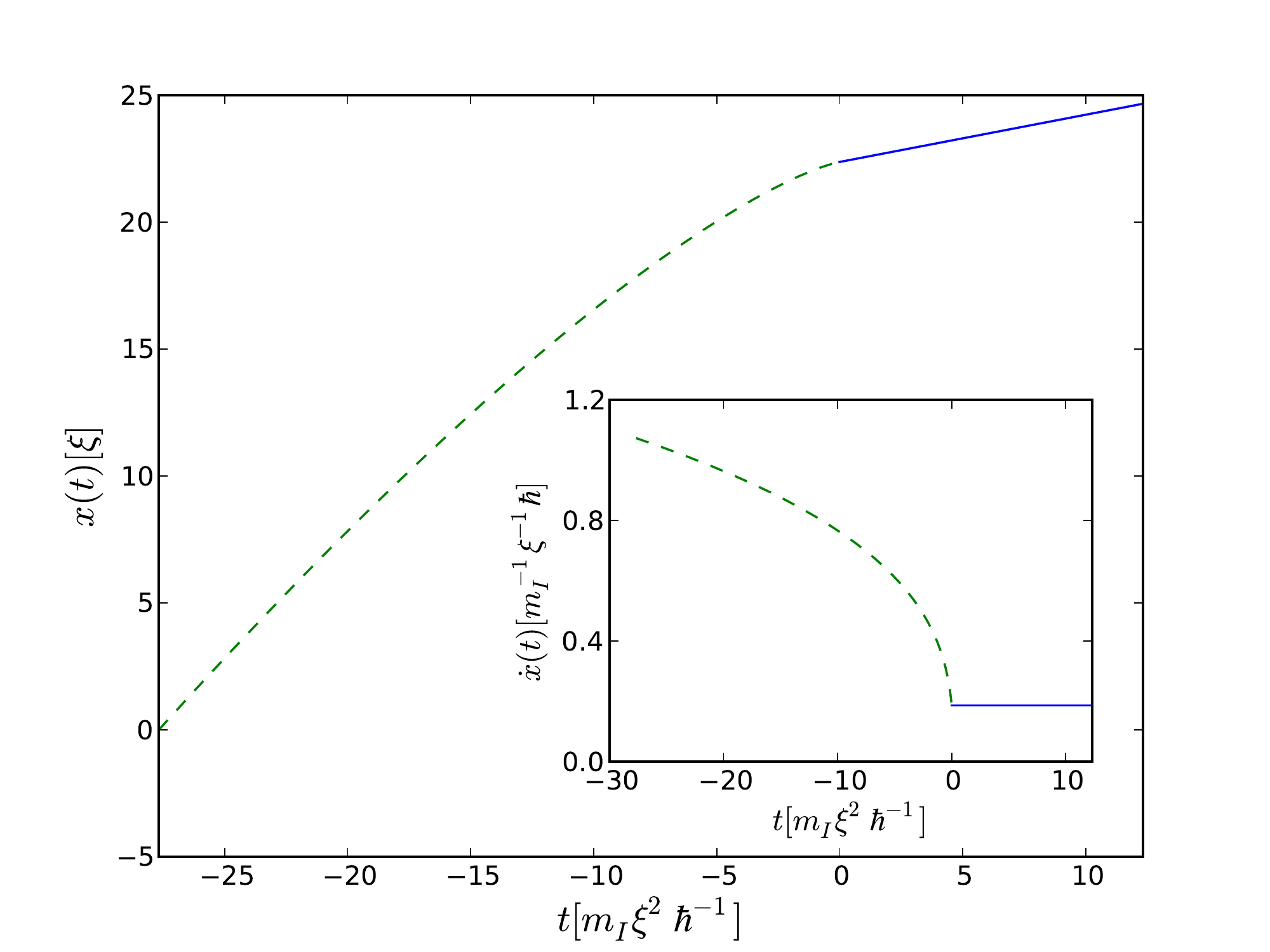}
  \caption{Asymptotic solution to the initial value problem with $v_0=4c$.  The main graph shows the evolution of the coordinate. The velocity (inset)
    decays according to a power law with exponent $1/3$, until it reaches a value of $c$, after which it
    remains constant and the particle moves freely. The time variable is gauged in such a way that 
the transition from the supersonic to the subsonic regime occurs at $t=0$.}
  \label{fig:asymptoticsolution}
\end{figure}

Especially the deceleration process of an initially supersonic impurity shows up interesting
features. As long as $v(t) > c$ the particle slows down and just above the threshold $c$ the
velocity decrease follows a power law \footnote{This power law is different for systems of different
  dimensionalities.} with exponent $1/3$, see Fig.~\ref{fig:asymptoticsolution}. After $v$ drops
below $c$ it does not change any more. From the qualitative point of view this behaviour is to be
expected, being a manifestation of superfluidity in the BEC. However, the change of velocity is
smooth in previous perturbative studies \cite{PhysLettA.282.421} as opposed to our result, in which
the time derivative of $v$ shows a jump. Furthermore, the emergence of the critical velocity $c$ is
clearly a consequence of the nonlinear coupling: The long time behaviour of the corresponding EOM in
which $\vb{r}=0$ is just that of a free particle, independent of its velocity, as can easily be
shown using Laplace transform techniques.

As we have already mentioned above, $k_c^{-1}$ can be connected to the conventional polaron
radius. To see this, we note that when we take into account quadratic fluctuations around the
classical path $\vb{x}(t)$ (see e.g.~\cite{JLowTempPhys.49.609,PhysRevB.30.6419}),
Eq.~(\ref{eq:eom}) becomes a generalized Langevin equation in which a noise term $\xi_i(t)$ has to
be added to the rhs. $\vb{\xi}(t)$ is a Gaussian stochastic process with average $\langle \xi_i(t)
\rangle = 0$ and state-dependent autocorrelation function
\begin{equation}
  \label{eq:noiseautocorrelation}
  \langle \xi_i(t) \xi_j(u) \rangle = \text{Re} \sum_\vb{k} k_i k_j \frac{\alpha k}{2} G_\vb{k}(t-u,
  \beta) e^{i \vb{k} \vb{r}},
\end{equation}
where $G_\vb{k}(t-u, \beta)$ is the real-time propagator for a boson of momentum $\vb{k}$ at inverse
temperature $\beta$. As required by the fluctuation-dissipation theorem, in the limit $\beta \to
\infty$, the rhs reduces to $\gamma_{ij}(t-u,\vb{r})/2$. When taking averages over realizations of
the noise of the full Langevin equation, $\xi_i(t)$ no longer explicitly occurs on the rhs of the
EOM, but we encounter averages of the form
\begin{equation}
  \label{eq:gaussiancharacteristicfunction}
  \langle e^{i \vb{k} \vb{r}} \rangle = e^{i \vb{k} \langle \vb{r} \rangle - k^2 (\langle \vb{r}^2
    \rangle - \langle \vb{r} \rangle^2 )/6},
\end{equation}
where we assumed that for every $t$, $u$, the random variable $\vb{r} = \vb{x}(t) - \vb{x}(u)$ can
be appropriately approximated by having Gaussian statistics. Therefore, from
Eq.~(\ref{eq:gaussiancharacteristicfunction}), we see that a natural cutoff for the $\vb{k}$-sums is
provided by the mean square of $\vb{r}$. In a stationary situation, we would expect that $\langle
\vb{r}^2 \rangle - \langle \vb{r} \rangle^2 \to \langle \vb{x}(t) \vb{x}(u) \rangle - \langle
\vb{x}(t) \rangle \langle \vb{x}(u) \rangle$, which for $u \to t$ can be considered to be a measure
for the squared polaron radius $r_{pol}$. Therefore, in this approximation
\begin{equation}
  \label{eq:cutoffpolaronradius}
  k_c \approx 6/r_{pol}^2 \, .
\end{equation}
For a coupling constant $\alpha$ of the order of unity, the polaron radius in the low temperature
limit has previously been calculated to be of the order of the healing length
\cite{PhysRevB.80.184504}, justifying both our use of the low-momentum approximations
(\ref{eq:epsilonk}) and of the semiclassical approximation.

We believe, that the phenomena described above can be very conveniently measured in state-of-the-art
experiments on ultracold fermion-boson mixtures. We envisage a dilute fermionic system being trapped
in an optical lattice, which, in turn, is immersed into a BEC as is done in
e.~g. Refs.~\cite{phdthesis_schuster,PhysRevA.85.042721}. 
Assuming the interwell fermion hopping to
be completely suppressed (for instance by making the lattice very deep) the lattice is then moved
through the BEC at constant velocity. The fermions are then subject to a drag force, which can be
measured as a population imbalance of the low-lying fermion states in the minimum of the potential
well by a number of well established experimental techniques.
%means of the Brillouin zone mapping technique \cite{PhysRevLett.94.080403}. 
Alternatively,
after the transients die out, one can switch off the trapping potentials and infer the force from the
velocity distributions of the freed particles.

\begin{figure}
  \centering
  \includegraphics[width=\columnwidth]{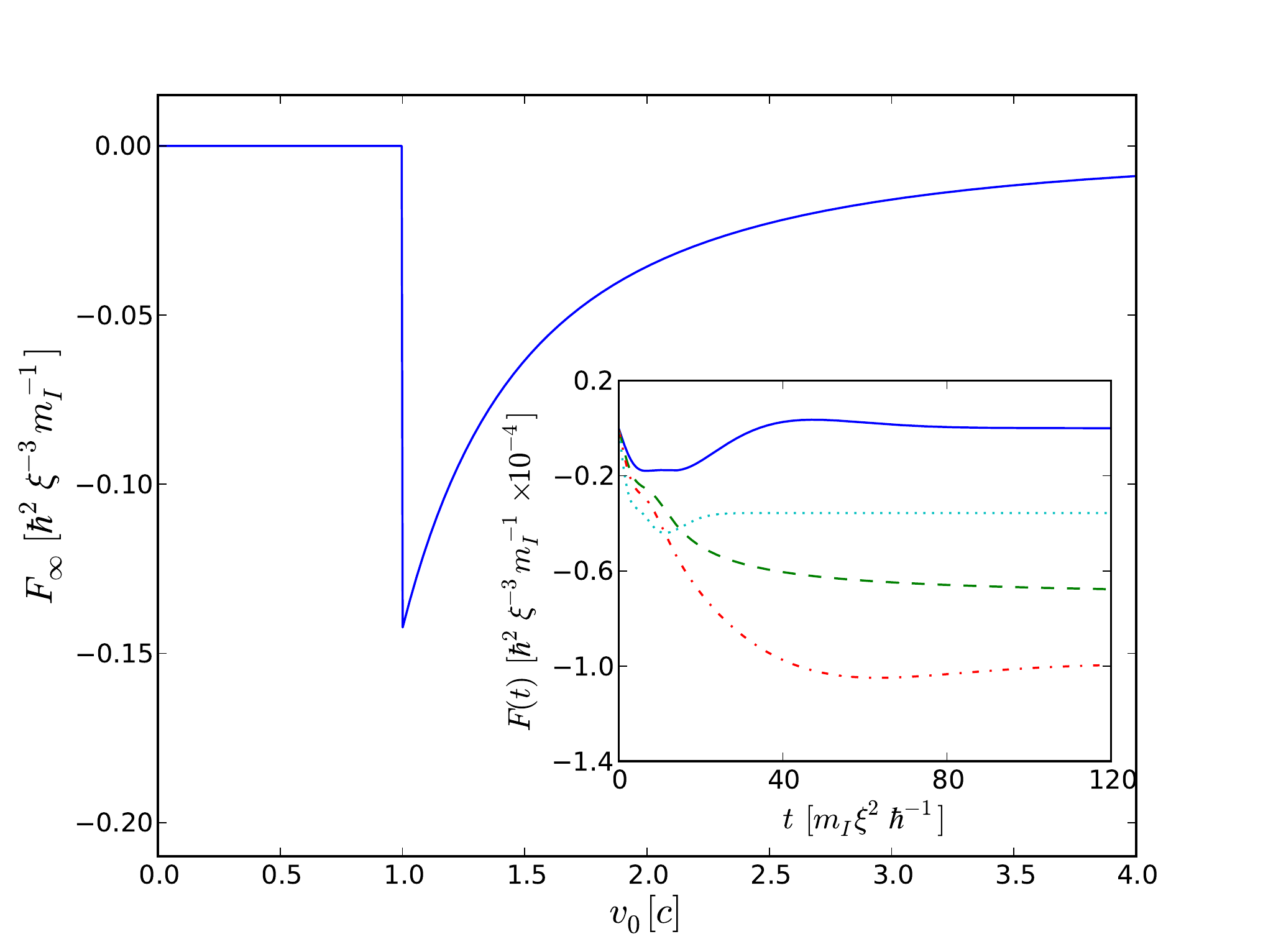}
  \caption{Asymptotic behavior of the drag force for $t \to \infty$ as a function of the drag velocity $v_0$. In
    the subsonic regime, it is zero as expected. It jumps to finite value just above $v_0=c$,
    subsequently decaying like $1/v_0^2$. Inset: drag force as a function of time for drag velocites
  $v_0=0.7c$ (solid line), $v_0=c$ (dashed line), $v_0=1.2c$ (dashed-dotted line) and $v_0=2c$
  (dotted line).}
  \label{fig:dragforce}
\end{figure}

From the theoretical point of view such an experiment is very convenient since the particle
kinematics is fully known. Assuming the impurity/polaron to be a classical particle, the drag force
is exactly given by the rhs of Eq.~(\ref{eq:eom}) under the condition ${\bf r}(t,u) = {\bf v_0}
(t-u)$, where ${\bf v_0}$ is the drag velocity. The result of calculations is shown in the inset of
Fig.~(\ref{fig:dragforce}). While in the subsonic case the asymptotic value of the force is zero, in
the supersonic case some finite value is reached in the stationary situation \footnote{Due to a
  factorized system preparation at $t=0$ and resulting polaron cloud formation process the force at
  intermediate times is finite. It roughly corresponds to the situation in which the impurity is
  `shot' into the BEC from the outside.}. Very interesting is the behavior of the drag force at $t
\to \infty$ as a function of $v_0$, which is given by the rhs of Eq.~(\ref{eq:longtimeeom}), see
Fig.~\ref{fig:dragforce}. It follows a $1/v_0^2$ law just beyond the sound velocity with a finite
value just above it. This is very different from the perturbative result $\sim v_0^4$ of the field
theoretical computation, see e.g.~\cite{PhysLettA.282.421}.  Since a mechanical force can be
interpreted as the amount of energy radiated per unit length, it is in fact more instructive to
compare this to the famous result for Cherenkov radiation in classical electrodynamics by Frank and
Tamm \cite{franktamm}, where this quantity scales as $1-c^2/v_0^2$, $c$ being the velocity of light in
a medium. Although in our case we have the same dependence on $1/v_0^2$ which seems to be a feature
of classical dynamics, the difference is that in the BEC case, we obtain an asymptotic decoupling
from the bath for high velocities as well as the discontinuity at $v_0=c$. 
The latter feature is plausible, it is clearly a signature of the nonlinear polaron coupling
mechanism since the discontinuity disappears if the coupling is linearized.

%Both of these features
%are plausible: The latter can be compared to a quantum phase transition if the velocity is
%interpreted as an order parameter, whereas the former seems to be an indication of asymptotic
%freedom of the underlying field theory. 

It would be very interesting to observe these effects in experiments. One possible realization could
be ${}^6\text{Li}$ fermionic impurities immersed into a BEC of ${}^{23}\text{Na}$
\cite{PhysRevB.80.184504}. In this setup the typical values for the force evaluate to $F_\infty
\approx 10^{-24}\,\text{N}$ in conventional units. This corresponds to an impurity acceleration of
the order of $10^{2} \, \text{m} / \text{s}^2$ and therefore seems to be well within reach of
experimental observation \cite{PhysRevA.85.042721,PhysRevB.80.184504}. As discussed above it might
be advantageous to replace the usually used fermion impurities by heavier ones, or even bosonic ones
like ${}^{85}\text{Rb}$. In this case the force would be of the order of $10^{-25}\,\text{N}$.

To conclude,  by means of the semiclassical approximation and its extensions we have investigated the dynamics of an impurity in a BEC and interacting with its Bogolyubov modes. It turns out to crucially depend on the velocity of the wave packet and is completely different in sub- and supersonic regimes. The resulting mechanical forces can be estimated and we expect them to be measurable in the upcoming experiments.

The authors would like to thank S.~Maier, M.~Oberthaler, T.~Rentrop, T.~Schuster, R.~Scelle and A.~Trautmann for many enlightening discussions. The authors are supported by the Centre of Quantum Dynamics and HGSFP of the University of Heidelberg.

\bibliography{Semiclassical_BEC_polaron.bib}

\end{document}